\documentclass[aps,prb,twocolumn,showpacs,superscriptaddress,floatfix,10pt]{revtex4-2}
\usepackage{amsmath,amssymb,amsfonts,stmaryrd,wasysym,graphicx,multirow,color,textcomp,subfigure}
\usepackage{url}
\usepackage[colorlinks=true,citecolor=blue,urlcolor=blue]{hyperref}
\usepackage{romannum}

\normalfont
\def\be{\begin{equation}}
	\def\ee{\end{equation}}
\def\bea{\begin{eqnarray}}
	\def\eea{\end{eqnarray}}

\newcommand{\cyan}[1]{{\color{cyan}{\em {#1}}}}

\begin{document}
	\title{Disorder-driven Phase Transitions of Second-order Non-Hermitian Skin Effects}
	
\author{Kyoung-Min Kim}
\affiliation{Center of Theoretical Physics of Complex Systems, Institute for Basic Science (IBS) Daejeon 34126, Republic of Korea}
\author{Moon Jip Park}
\email{moonjippark@ibs.re.kr}
\affiliation{Center of Theoretical Physics of Complex Systems, Institute for Basic Science (IBS) Daejeon 34126, Republic of Korea}

	\begin{abstract}
		Non-Hermitian skin effect exhibits the collapse of the extended bulk modes into the extensive number of localized boundary states in open boundary conditions. Here we demonstrate the disorder-driven phase transition of the trivial non-Hermitian system to the higher-order non-Hermitian skin effect phase. In contrast to the clean systems, the disorder-induced boundary modes form an arc in the complex energy plane, which is the manifestation of the disorder-driven dynamical phase transition. At the phase transition, the localized corner modes and bulk modes characterized by trivial Hamiltonian coexist within the single-band but are separated in the complex energy plane. This behavior is analogous to the mobility edge phenomena in the disordered Hermitian systems. Using effective medium theory and numerical diagonalizations, we provide a systematic characterization of the disorder-driven phase transitions.
	\end{abstract}
	
	\maketitle
	
\cyan{Introduction} - The unique physical behaviors of non-Hermitian mechanics appear in various open systems such as optics\cite{Longhi:20,PhysRevResearch.2.013280,Feng2017,Mirieaar7709,ozdemir2019}, electrical circuits\cite{Li2020,PhysRevResearch.2.023265,Helbig2020,zou2021observation}, mechanical systems\cite{zhang2021observation,Ghatak29561,Brandenbourger2019,PhysRevResearch.2.023173,PhysRevLett.125.118001}, open quantum systems\cite{PhysRevLett.123.170401,PhysRevLett.124.250402,PhysRevLett.124.040401}, and correlated quantum systems with a finite lifetime\cite{kozii2017nonhermitian,PhysRevB.99.201107,PhysRevLett.121.026403,PhysRevB.99.041116,PhysRevB.98.035141,PhysRevB.98.155430}. The non-Hermitian skin effect(NHSE) is an exotic example of non-Hermitian mechanics, in which the bulk mode shows a dramatic difference depending on the boundary condition\cite{PhysRevLett.121.086803,PhysRevLett.123.246801,PhysRevB.103.L140201,PhysRevB.100.054301,PhysRevB.103.144202,PhysRevX.8.031079,PhysRevB.103.L140201}. For example, Hatano-Nelson model demonstrates the one-dimensional NHSE, which all bulk modes in periodic boundary conditions collapse into the localized mode in the open boundary condition. The one-dimensional NHSE is characterized by the topological winding number of the energy spectrum in the complex energy plane
\cite{PhysRevLett.121.086803,PhysRevLett.77.570,PhysRevLett.121.026808,PhysRevLett.123.066404,PhysRevB.99.245116}. In addition, the second-order NHSE has been recently proposed\cite{PhysRevB.102.205118,PhysRevB.102.241202,PhysRevB.103.045420}. In these systems, the non-Hermitian topology of the $d$-dimensional bulk realizes ($d-2$)-dimensional boundary modes on the corner, which has a close analogy with the Hermitian higher-order topological insulator phases.

Although the NHSE exhibits similar boundary modes with the Hermitian topological insulator phases, the crucial difference lies in the number of the boundary modes. In non-Hermitian systems, the number of the topological boundary mode is an extensive quantity, which proportionally grows with the system size. For example, in the first-order NHSE, $O(L^{d})$ number of the boundary modes emerges in $d$-dimensional sized system where $L$ is the length of the system in each direction. Similarly in the second-order NHSE, $O(L^{d-1})$ boundary modes occur\cite{PhysRevB.102.205118}. The extensiveness of the edge mode is the hallmark of the non-Hermitian system that is directly contrast with the Hermitian topological insulators.

In this work, we study the disorder-driven phase transition of the second-order NHSE. Although the disorder-induced topological phase transition has been extensively studied in the Hermitian topological systems\cite{PhysRevLett.102.136806,PhysRevLett.103.196805,PhysRevLett.110.236803,PhysRevLett.114.056801,PhysRevB.85.155138,PhysRevLett.105.216601,PhysRevB.95.094201,PhysRevLett.126.206404,PhysRevLett.125.166801}, we show that the extensiveness of the boundary modes plays a crucial role, and we newly discover the dynamical phase transition of the NHSE. At the phase transition, we observe a novel mobility edge phenomena, which the bulk energy spectrum is separated into the trivial bulk modes and the NHSE bulk modes characterized by second-order NHSE Hamiltonian. The physical manifestation of this dynamical phase transition is the NHSE corner modes, which form an arc in the complex energy plane. As a result, we find that a single band spectrum shows the coexistence of the NHSE bulk modes and the trivial bulk modes. Our work reveals rich physical behaviors of the disordered NHSE in non-Hermitian systems.

\begin{figure}[t!]
	\centering\includegraphics[width=0.5\textwidth]{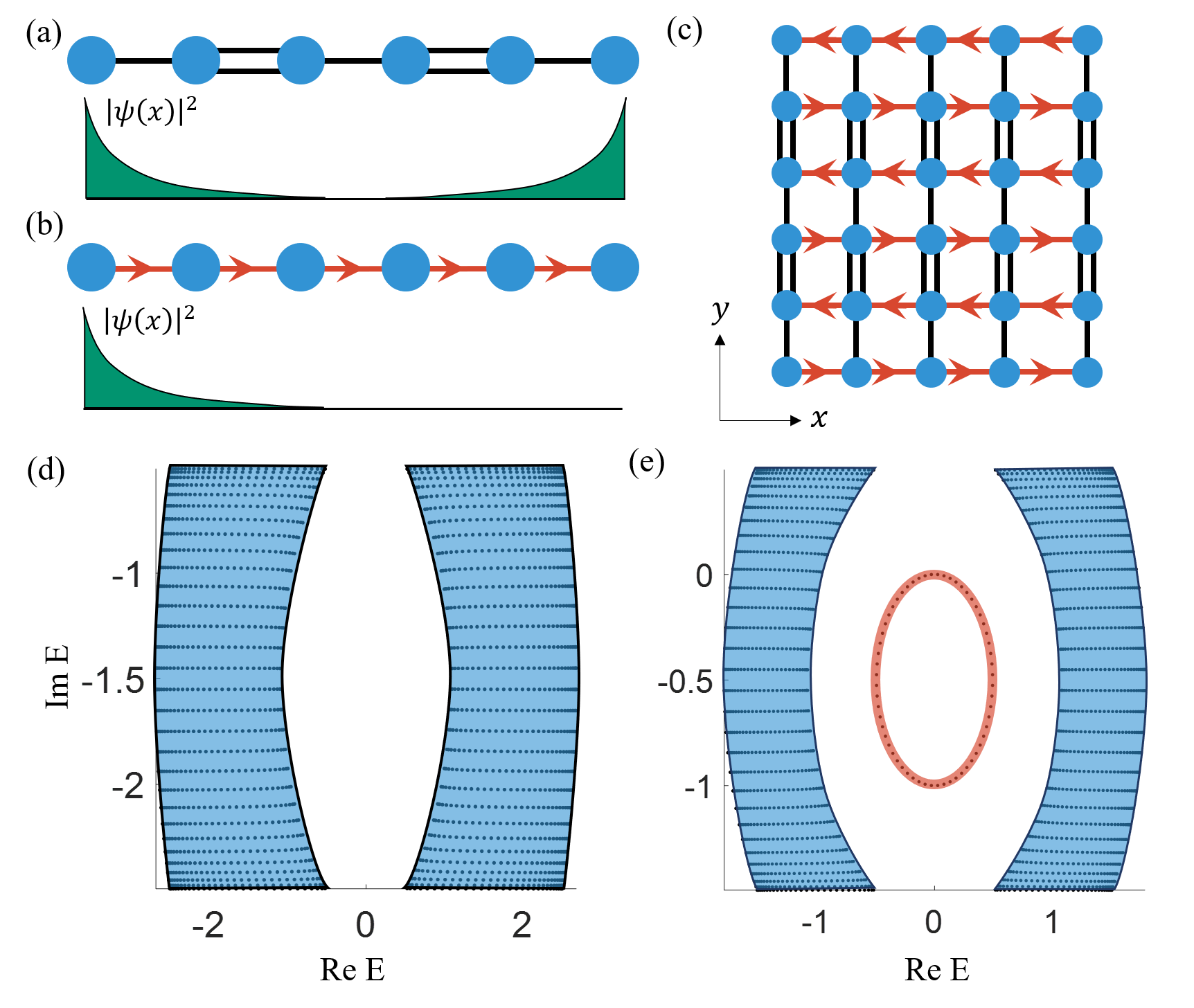}
	\caption{ (a)-(c) Schematic illustration of (a) the one-dimensional SSH chain, (b) Hatano-Nelson model, and (c) the higher-order NHSE. While the SSH chain with the dimerized hoppings hosts the pair of the localized boundary modes, the Hatano-Nelson model with the asymmetric hopping (colored by the red arrow) exhibits the collapse of bulk modes into the localized modes at only one end. The model of the Higher-order NHSE is constructed by the asymmetric non-Hermitian hopping term in $\hat{x}-$direction and the dimerized hopping in $\hat{y}$-direction. The physical manifestation of the higher-order NHSE is $O(L)$ numbers of the corner modes. (d)-(e) The typical band structure of the second-order NHSE in (d) trivial and (e) non-trivial regime. The blue surface and red line represents the bulk modes and the NHSE corner modes respectively. } 
	\label{fig:1} 
\end{figure}

\cyan{Higher-order skin effect in clean limit} - 
To construct the model of the second-order NHSE, we consider the following dual relation of the non-Hermitian Hamiltonian, $H_{\textrm{NH}}(\mathbf{k})$ and the extended Hermitian Hamiltonian $H_{\textrm{BBH}}(\mathbf{k})$ as\cite{PhysRevB.102.205118},
 
 \bea
 H_{\textrm{BBH}}(\mathbf{k})=
 \begin{pmatrix}
 	0 & H_{\textrm{NH}}(\mathbf{k}) \\
 	H_{\textrm{NH}}^\dagger(\mathbf{k})& 0
 \end{pmatrix},
 \eea
where $H_{\textrm{BBH}}(\mathbf{k})$ is the Bloch Hamiltonian of the Benacazar-Bernevig-Hughes(BBH) model\cite{Benalcazar61}, which shows the Hermitian higher-order topological insulator phase. The non-Hermitian dual Hamiltonian, $H_{\textrm{NH}}(\mathbf{k})$, is explicitly given as,
\bea
H_\textrm{NH}(\mathbf{k})= -i (\gamma+\lambda \cos k_x)+(\gamma+\lambda \cos k_y)\sigma_y
\nonumber
\\+\lambda(\sin k_x\sigma_z+\sin k_y\sigma_x),
\label{Eq:model}
\eea
where $\sigma_i$ is $i$-th Pauli matrices. In real space, this model forms the Su-Schrieffer-Heeger(SSH) like dimerized chain along $k_y$ direction and the Hatano-Nelson like asymmetric hopping terms along $k_x$ direction (See Fig. \ref{fig:1}). The bulk energy spectrum is given as, $E(\mathbf{k})=\pm\sqrt{(\gamma+\lambda \cos k_y)^2+\lambda^2(\sin^2 k_x+ \sin^2 k_y)}-i(\gamma+\lambda \cos k_x)$, which exhibits the line gap along the imaginary axis in the complex energy plane ($\textrm{Re} z=0$). The line gap closes when $|\gamma /\lambda|=1$ accompanying the topological phase transition of the second-order NHSE. The second-order NHSE occurs when $|\gamma /\lambda|<1$, and the physical manifestation is the emergence of the localized corner modes in the open boundary condition (Red lines in Fig. \ref{fig:1} (d)). In contrast to the Hermitian second-order topological insulator, the number of the corner modes grows in order of $O(L)$, where $L$ is the system length. 

$C_4$-rotational symmetry of the BBH model topologically protects the second-order NHSE in the non-Hermitian dual Hamiltonian as well as the higher-order topological insulator phase. The condition of the $C_4$-rotational symmetry in the Hermitian Hamiltonian is translated into the non-Hermitian Hamiltonian as, $-i\sigma_y H^\dagger_{\textrm{NH}}(k_x,k_y)=H_{\textrm{NH}}(-k_y,k_x)$. We can introduce the $C_4$-rotational symmetry protected winding number under the defect classification\cite{PhysRevB.81.134508,RevModPhys.88.035005}. To do so, we introduce the additional auxiliary parameter $t\in [0,1]$, and consider the adiabatic deformation of the non-Hermitian Hamiltonian into the trivial atomic insulator as, $H(\mathbf{k},t=0)=H_\textrm{NH}(\mathbf{k})$ and $H(\mathbf{k},t=1)=\sigma_x$. During the adiabatic deformation, the non-Hermitian Hamiltonian can be singular-value decomposed as, $H(\mathbf{k},t)= U_\mathbf{k}^\dagger D_\mathbf{k} V_\mathbf{k}$, which allows to define a unitary matrix, $q_\mathbf{k}\equiv U_\mathbf{k}^\dagger V_\mathbf{k}=\sum_n e^{i\lambda_{n}(\mathbf{k})}|n(\mathbf{k}) \rangle \langle n(\mathbf{k})|$, and the corresponding phase  $\lambda_{n}(\mathbf{k})$. Using the unitary matrix, $q_\mathbf{k}$, we can define $\mathbb{Z}_2$-valued three-dimensional winding number of $\lambda_{n}(\mathbf{k})$ as\cite{PhysRevB.82.115120},
\bea
\mathcal{W}=\frac{1}{24\pi^2}\int^1_0 dt \int d^2 \mathbf{k} \epsilon^{ijk} \textrm{Tr} [q_{\mathbf{k}}^\dagger \partial_i q_{\mathbf{k}} q_{\mathbf{k}}^\dagger \partial_j q_{\mathbf{k}} q_{\mathbf{k}}^\dagger \partial_k q_{\mathbf{k}}].
\eea
It is shown that $C_4$-rotational symmetry quantizes the winding number as it takes non-trivial (trivial) value, $\mathcal{W}=1/2 (0)$ if $|\gamma /\lambda|<1$($|\gamma /\lambda|>1$) \cite{PhysRevB.102.205118}.

\cyan{Disorder induced corner arc modes} - After establishing the second-order NHSE in the clean limit, we now consider the addition of the on-site random disorder in the Hamiltonian in Eq. \eqref{Eq:model}. Motivated by the $C_{4}$-rotational symmetry we first analyze the particular type of the onsite disorder, $V_{\textrm{dis}}=\sum_i \omega_i (I_2-i\sigma_y) c^\dagger_i c_i $. Here $c_i$ is the annihilation operator in $i$-th site, and $w_i$ is the uniformly distributed random number within the window of $w_i\in [-W/2,W/2]$. The introduction of the random disorder term immediately breaks the translational symmetry of the systems. However, the effective Bloch Hamiltonian can be derived by averaging many disorder configurations until it restores the translational symmetry. By performing the disorder averaging, we first numerically compute the density of states(DOS), $P(z)$, in the complex energy plane as\cite{PhysRevLett.79.491},

\bea
P(z)=\frac{1}{\pi N} \lim_{\eta \rightarrow 0} \langle \sum_{i} \frac{\eta^2}{(\textrm{Re}z-\epsilon'_i)^2+(\textrm{Im}z-\epsilon''_i)^2+\eta^2} \rangle ,
\nonumber
\\
\eea
where $N$ is the total number of the states. $\langle ... \rangle$ indicates the averaging over distinct disorder configurations. $\epsilon_i=\epsilon'_i+i\epsilon''_i$ is $i$-th complex eigenenergy. $\eta$ is the infinitesimal real number that introduces the broadening of the quasiparticle peaks in the complex energy plane.

\begin{figure}[t!]
	\centering\includegraphics[width=0.5\textwidth]{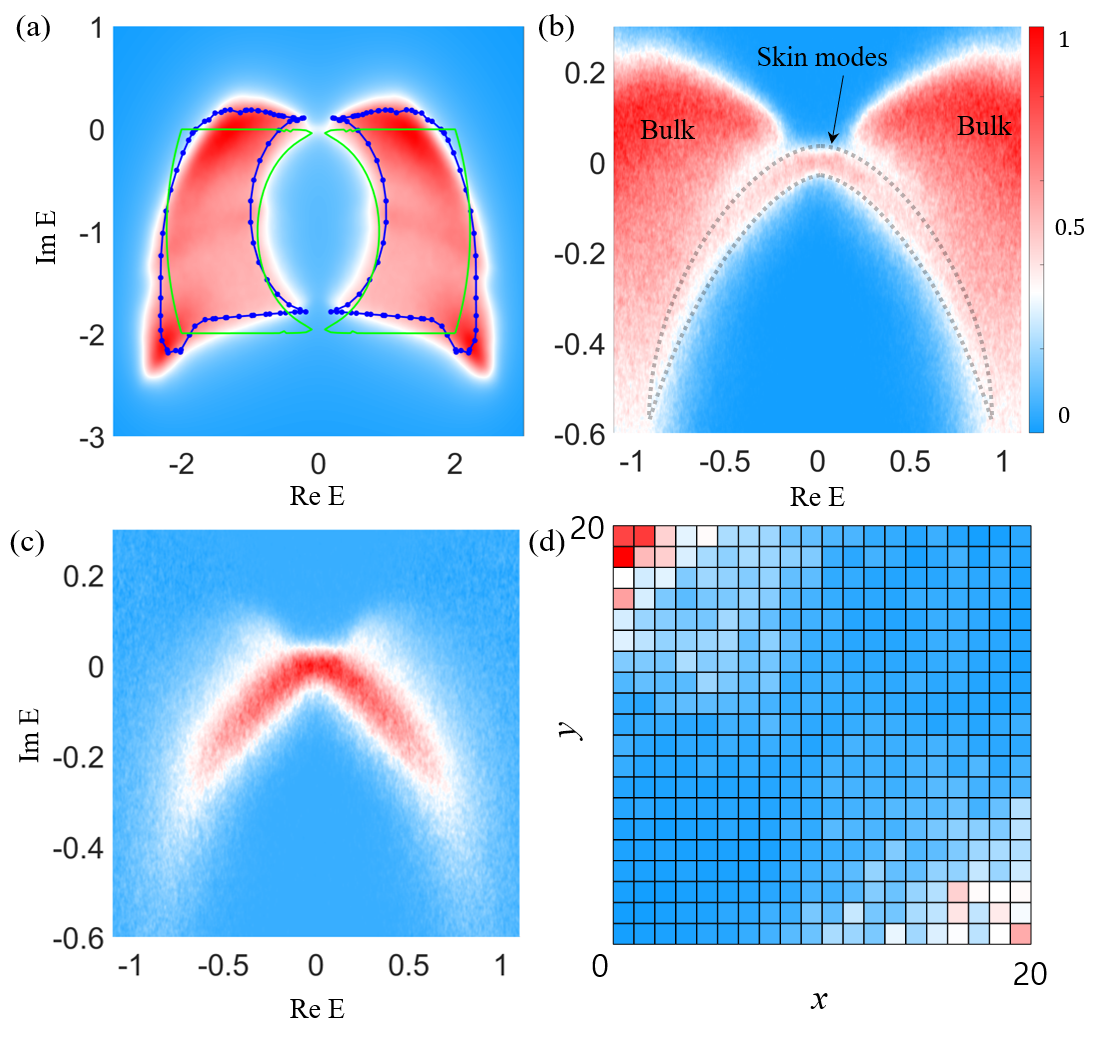}
	\caption{ (a) The DOS in the complex energy plane. The surface plot indicates the DOS in the presence of the disorder obtained by the brute-force numerical diagonalization. The green solid line is the band structure in the clean limit. The blue dotted line indicates the deformed band structure derived using the SCBA. (b) The disorder-induced corner modes in the open boundary condition. The corner modes form an arc in the complex energy plane (dashed lines). (c) The expectation value of the localization at the corner. The region with the corner arc modes has a non-zero value, indicating the localization at the corner. (d) typical wave function profile of the disorder-induced corner modes in real space.}
	\label{fig:2} 
\end{figure}

Fig. \ref{fig:2} shows the density of states at the critical point of the topological phase transition, ($\gamma/\lambda=1$) in the presence of the disorder ($W=2$). We observe a clear deviation of the disordered density of states, compared to the band structure in the clean limit  (red surface and green solid lines in Fig. \ref{fig:2} (a)). The deformation of the effective band structure has strong energy dependence in the complex energy plane. For example, in the upper half part of the band structure, the bandwidth along the real axis suppresses, while the lower half part shows the extension of the bandwidth. This contrasting tendency in the deformation of the band structure indicates the strong energy-dependent renormalization due to the disorder. As we rigorously show in the next section using the effective medium theory, the overall band deformation can be explained by the renormalization of the topological mass, $\gamma $, in Eq. \eqref{Eq:model}. Furthermore, the renormalization of the topological mass drives the disorder-induced second-order NHSE in the trivial non-Hermitian systems.

In addition, as we take open boundary condition along with both $\hat{x}-$ and $\hat{y}-$ directions, we find the emergence of the NHSE corner modes at the upper half-plane of the band structure (dashed line in Fig. \ref{fig:2} (b)). This corner mode is induced by the disorder and appears as the form of the arc, in which the tip of the arc is absorbed into the bulk states. This result is rather unusual since the NHSE corner modes only occur as the closed ring in the clean limit (dashed line in Fig. \ref{fig:1} (d)). This result is the signature of the novel dynamical phase transition of the second-order NHSE, where the upper half-plane of the complex energy becomes topological($\textrm{Im} z>-\gamma$) and the lower half-plane is trivial ($\textrm{Im} z<-\gamma$). As a result, the horizontal line with $\textrm{Im} z\approx -\gamma$ in the complex energy plane separates the topologically non-trivial modes and the trivial modes in the complex energy plane. This complex energy-dependent phase transition of the boundary modes has not been observed in the Hermitian systems. As we show in the next section, it is uniquely observed to the non-Hermitian systems, where an extensive number of boundary modes exist in the complex energy plane.

\cyan{Effective medium approximation}- The disorder-induced renormalization of the band structure can be better understood by considering the effective medium approximation. To do so, we introduce the disorder-averaged effective Green function, which is given as,
\bea
G_{\textrm{eff}}(z)\equiv  \frac{1}{z-H_\textrm{eff}(z)}=\langle \frac{1}{z-H_\textrm{NH}-H_{\textrm{dis}}} \rangle.
\eea
where $H_{\textrm{eff}}(z)\equiv H_{\textrm{NH}}+\Sigma (z)$ represents the effective Hamiltonian derived by averaging Green functions over the different disorder configurations.
If the number of the disorder configuration is sufficiently large, the effective Hamiltonian $H_{\textrm{eff}}$ recovers the translational symmetry, and the effective Bloch Hamiltonian can be defined. The modification of the band structure due to the disorder can be expressed as the self-energy correction, $\Sigma (z)$. We derive the self-energy correction, using the standard Born approximation. In the leading order, the Born approximation gives rise to the following self-consistent equation of the self-energy corrections\cite{PhysRevLett.79.491,PhysRevLett.80.2897}.
\bea
\nonumber
\Sigma(z)=\frac{W^2}{12N}\sum_\mathbf{k} (I-i\sigma_y) \frac{1}{z-H_\textrm{NH}(\mathbf{k})-\Sigma(z)} (I-i\sigma_y) .
\\
\label{eq:born}
\eea
where $N$ is the number of the total sites in the systems. Since the clean Hamiltonian in Eq. \eqref{Eq:model} proportional to $\sigma_x$ and $\sigma_z$ terms has the odd momentum dependence, the self-energy terms with $\sigma_x$ and $\sigma_y$ vanish during the momentum sum in Eq. \eqref{eq:born}. Therefore, we only need to consider the self-energy correction in $\textrm{I}_2$ and $\sigma_y$ terms as,
\bea
\Sigma(z)=\Sigma_0(z) \textrm{I}_2+\Sigma_y(z) \sigma_y.
\eea
(See supplementary materials for the detailed calculations). The self-energy corrections correspond to the renormalization of the topological mass as,
$
\bar{\gamma}_0(z)=\gamma+\Sigma_0(z), 
\quad \bar{\gamma}_y(z)=\gamma+\Sigma_y(z),
$
where $\bar{\gamma}_0$ and $\bar{\gamma}_y$ represent the renormalized $\gamma$ term in Eq. \eqref{Eq:model} that is proportional to $\textrm{I}_2$ and $\sigma_y$ respectively. Blue dotted lines in Fig. \ref{fig:2} shows the numerically calculated band structure obtained using the self-consistent Born approximation(SCBA) of Eq. \eqref{eq:born}. We find that the SCBA and the disordered band structure obtained by the numerical diagonalization agree very well.

\begin{figure}[b!]
	\centering\includegraphics[width=0.5\textwidth]{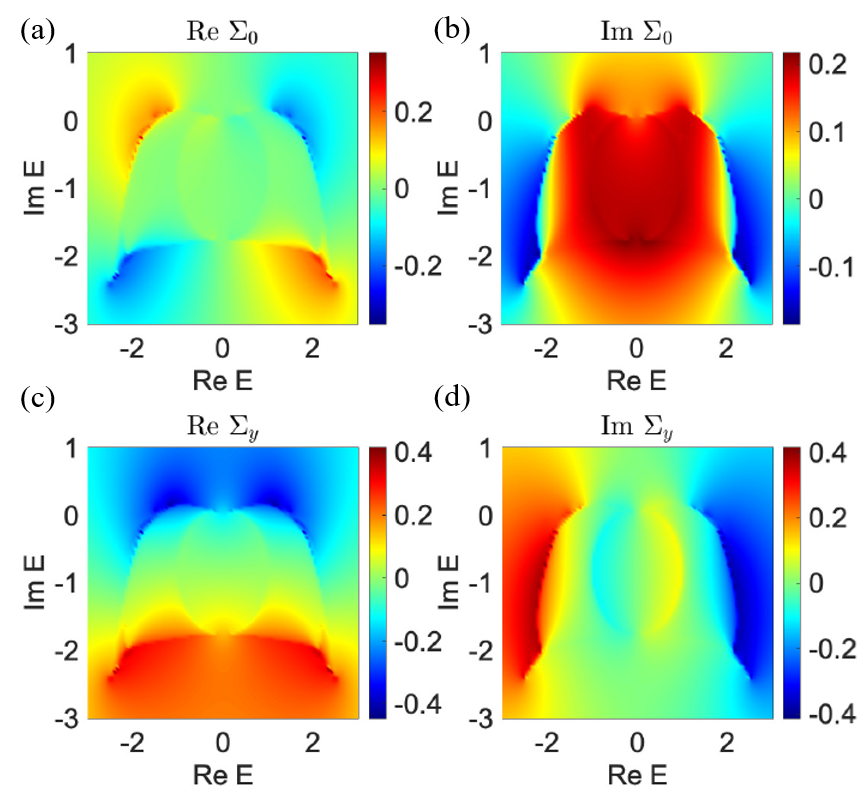}
	\caption{ The self-energy correction calculated using the SCBA at $\gamma/\lambda=1$. Figure (a)-(b) and (c)-(d) represent the correction in $\Sigma_0$ and $\Sigma_y$ respectively. The energy dependence of the self-energy induces the dynamical phase transition of the second-order NHSE.}
	\label{fig:3} 
\end{figure}

We now analyze the self-energy term in more detail. Fig. \ref{fig:3} shows the calculated self-energy corrections in the complex energy plane at the critical point, $\gamma/\lambda=1$. Since $\Sigma_0(z)$ only shifts the overall energy of the Bloch Hamiltonian. We only need to consider the correction in $\Sigma_y(z)$ to capture the change in the topology. In particular, we find that the correction in $\textrm{Re}\Sigma_y$ term changes the sign in the complex energy plane as, $\textrm{Re}\Sigma_y(z)<0$ $(>0)$ when $\textrm{Re}z>0$ $(<0)$. As a result, the effective Hamiltonian along the imaginary axis, $H_{\textrm{eff}}(\textrm{Re}z=0)$, is characterized by the topologically non-trivial mass term when $\textrm{Im}z>-\gamma$, while it becomes trivial when $\textrm{Im}z<-\gamma$. This energy-dependent correction in  $\bar{\gamma}_y(z)$ gives rise to the asymmetric bulk band shape and the emergence of the NHSE skin effect at the upper-half plane of the complex energy. Finally, the disorder induces the dynamical phase transition of the higher-order NHSE. The physical manifestation is the arc of the NHSE corner modes that appears in the upper half-plane.

\begin{figure}[t!]
	\centering\includegraphics[width=0.5\textwidth]{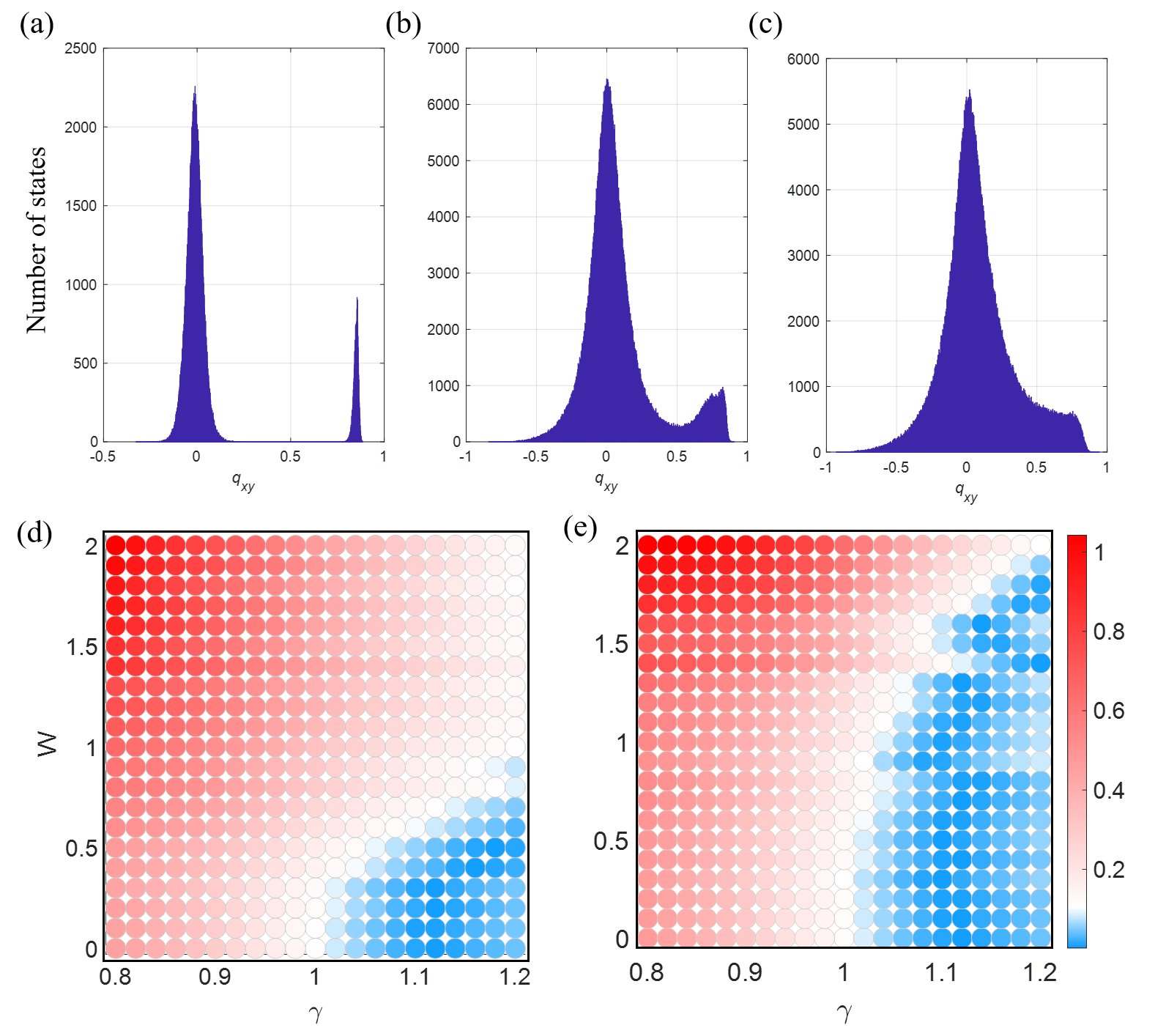}
	\caption{ Distribution of the quadrupole moment as a function of the disorder strength (a) $W=0.5$, (b) $W=1.5$, and (c) $W=2.0$. For moderate strength of the disorder, we find that the distribution of the corner modes and the extended bulk modes are well-separated. As the disorder strength further increases, the Anderson localization occurs. The trivial localized modes and the corner modes become incomparable in the distribution. (d)-(e) phase diagram showing the averaged $q_{xy}$ as a function of the topological mass, and the disorder strength for (d) the disorder, $V_{\textrm{dis}}$, and (e) the Anderson disorder. }
	\label{fig:4} 
\end{figure}

\cyan{Robustness of disorder-induced NHSE} - As the disorder strength further increases, we find that the quasiparticle peak broadens. Eventually, the spectrum of the corner boundary modes overlaps with the bulk modes. Moreover, Fig. \ref{fig:4} (a)-(c) shows the distribution of the quadrupole moment of each eigenstate, which is given as,
\bea
q_{xy}=\frac{1}{(L/2)^2}\langle  (x-\bar{x})(y-\bar{x}) \rangle,
\eea
where $q_{xy}=1 (0)$ corresponds to the perfectly localized states at the corner (extended states in the bulk). In moderate strength of the disorder, the two distinct peaks are observed, which separates the conventional bulk modes and the boundary modes from the NHSE. However, as the disorder strength increases up to $W\approx 2$(Fig. \ref{fig:4} (c)), the Anderson localization occurs. As the bulk modes are localized at the corner, the distribution of the bulk states overlaps with the corner modes. Eventually, the Anderson localized modes become incomparable to the NHSE modes. Finally, Fig. \ref{fig:4} (d) depicts the averaged value of $q_{xy}$ for whole states as a function of the disorder strength and $\gamma$. The general trend shows the increase of $q_{xy}$ near the critical point, signifying the disorder-induced NHSE phase transition. This trend continues until the effective medium theory fails to account for the broadened quasiparticle spectral function and the Anderson localization occurs. In the case of the Anderson type disorder, $V_{\textrm{dis-and}}=\sum_i \omega_i I_2 c^\dagger_i c_i $(Fig. \ref{fig:4} (e)), we also observed the similar disorder-driven topological phase transitions but in this case the role of $\Sigma_0$ and $\Sigma_y$ is reversed. (See supplementary material for the detailed calculation of the spectral function.)

\cyan{Discussions} - In conclusion, we study the disorder-driven phase transitions of the second-order NHSE. Using the numerical diagonalization, we find that the random disorder induces the phase transition of the second-order NHSE. This phase transition can be systematically understood using the effective medium theory. We have clarified that the phase transition is induced by the renormalization of the topological mass, which induces the second-order NHSE. Furthermore, the renormalization of the topological mass has a strong energy dependence, where the NHSE bulk modes and the trivial bulk modes coexist in the complex energy plane. The physical manifestation is the arc of the NHSE corner modes, that appears only at certain regions in the complex energy plane. Such dynamical phase transition is the unique feature of the non-Hermitian system that arises due to the extensiveness of the NHSE. This disorder-induced phase transition can be experimentally realized in tunable non-Hermitian systems such as topoelectric circuit experiments\cite{zou2021observation} and active matter systems\cite{palacios2020guided}.

	\acknowledgments
	
	M.J.P. and K-.M. Kim thank Hee Chul Park, Jung-Wan Ryu, Sungjong Woo, Jae-Ho Han, Chang-Hwan Yi, and Hyeoung Jun Lee for fruitful discussions.
	
	\bibliography{reference}
	
	\clearpage
	\pagebreak
	
	\renewcommand{\thesection}{\arabic{section}}
	\setcounter{section}{0}
	\renewcommand{\thefigure}{S\arabic{figure}}
	\setcounter{figure}{0}
	\renewcommand{\theequation}{S\arabic{equation}}
	\setcounter{equation}{0}
	
\begin{widetext}
\section{Supplementary Material}

\subsection{numerical diagonalization method}
\subsubsection{Order parameter of phase transition}
To obtain the phase diagram in Fig. \ref{fig:4} (d) and (e), we numerically compute the following order-parameter,
\begin{equation}
O = \frac{1}{N_{s}}\sum_{s} \frac{1}{N_{corner}} \sum_{n} q_{n}, \label{eq: order parameter}
\end{equation}
where $q_{n}$ measures the localization of the eigenstate $\left| n \right \rangle$ at the corner of the lattice as follows
\begin{equation}
q_{n} = \frac{1}{(L/2)^{2}}\left\langle n \right | (x-\bar{x})(y-\bar{y}) \left| n \right\rangle. \label{eq:quadrupole moment}
\end{equation}
Here, $L$ is the system size and $\bar{x}$ and $\bar{y}$ are the center position in the $x$ and the $y$ direction, respectively. $q_{n}  = 1$ or $-1$ if the wave function of $\left | n \right \rangle$ is completely localized at one of the four corners. A clean system is in a topological phase if there are corner states with nonzero values of $q_{n}$. Otherwise, it is in a trivial phase. To extend this identification to disordered systems, we sum $q_{n}$ over all eigenstates and take an average over distinct disorder configurations as shown in Eq. (\ref{eq: order parameter}). We note that $q_{n}$ of bulk states have random values in each disorder configuration so that the bulk state contribution would vanish in disorder average. The contribution from corner states, however, can survive in disorder average as follows
\begin{equation}
\lim_{N_{s}\rightarrow\infty} \bigg( \frac{1}{N_{s}}\sum_{s}\sum_{n \in ~ \textrm{bulk}} q_{n} + \frac{1}{N_{s}}\sum_{s}\sum_{n \in ~ \textrm{corner}} q_{n} \bigg) \rightarrow  \lim_{N_{s}\rightarrow\infty}  \frac{1}{N_{s}}\sum_{s}\sum_{n \in ~ \textrm{corner}} q_{n}.
\end{equation}
Finally, we normalize $O$ by dividing it with the number of corner states $N_{corner}=\sqrt{2N}$ such that $O=1$ if $q_{n}=1$ for all corner states where $N$ is the number of lattice sites.

Figure \ref{figs:figs1} (a) shows the evolution of the order parameter with the sample size $N_s$. With more than one hundred samples or so, the order parameter reach plateaus where the estimates become reliable. This observation leads us to take $N_s$ from $100$ to $400$ in our computation. Figure \ref{figs:figs1} (b) shows the result with $\lambda=1$ and $\gamma=1$. The relative standard errors are less than $0.05$ for most of data points so it seems that the disorder averaged results are quite reliable. 
 
\begin{figure}[t!]
	\centering\includegraphics[width=\textwidth]{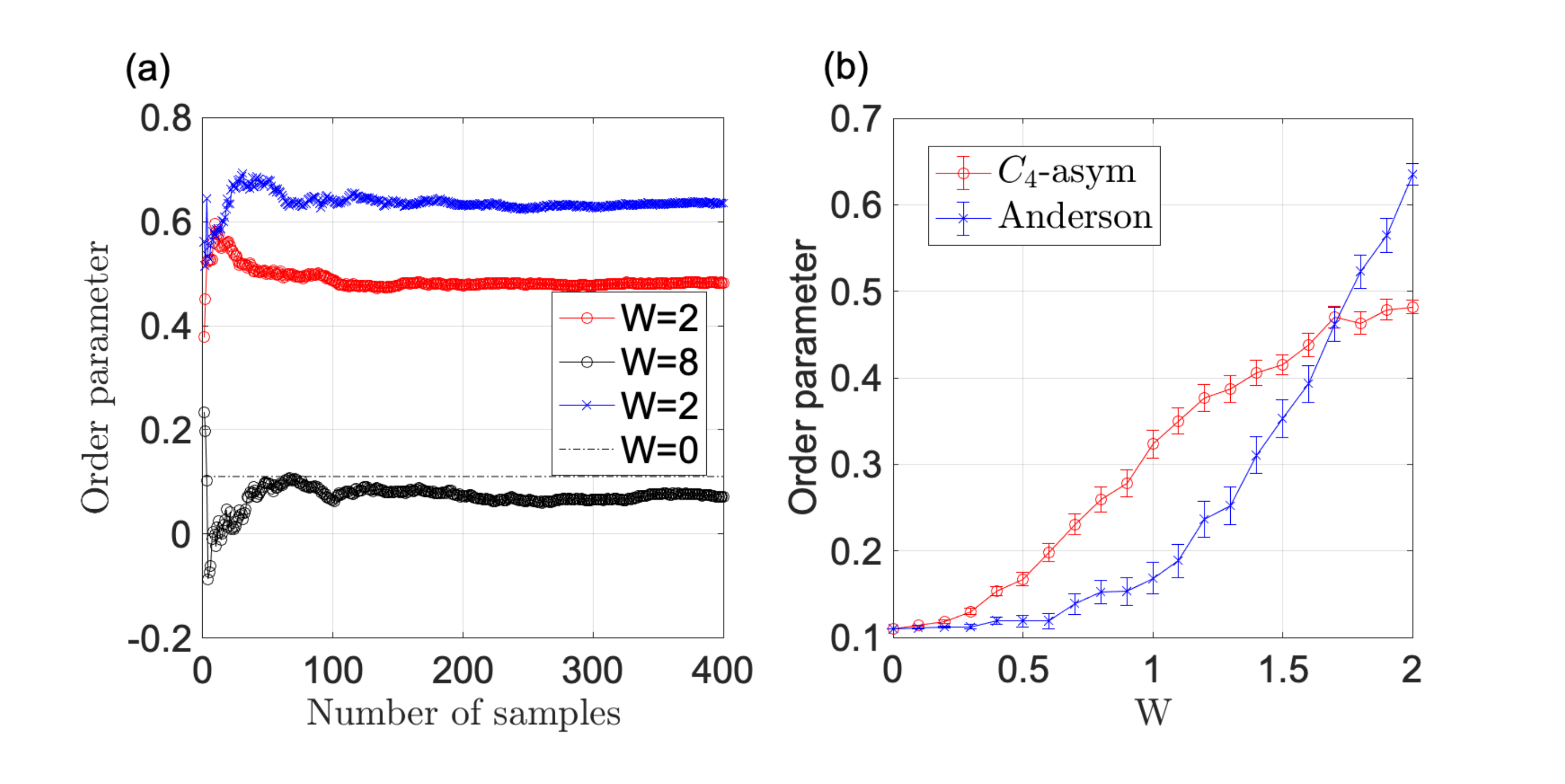}
	\caption{(a) Evolution of order parameter as the number of samples of distinct disorder configurations is increased. The red and the black markers are for $C_{4}$-asymmetric disorder and a blue one for Anderson disorder. (b) Order parameter as a function of disorder strength. The error bars represent standard errors in disorder average. In all plots, $\lambda=1$ and $\gamma=1$ are used.  }
	\label{figs:figs1} 
\end{figure}

\subsubsection{Density of states with numerical diagonalization}
To compute the disordered density of states in Fig. \ref{fig:2} (a) and (b), we utilized the following expression
\begin{equation}
P(z) = \frac{1}{N_{s}}\sum_{s}\frac{1}{N}\sum_{n} p(z,\epsilon_n),
\end{equation}
where $p(z,\epsilon_{n})$ is a Lorentzian function given by
\begin{equation}
p(z,\epsilon_{n}) = \frac{1}{\pi} \frac{\eta}{(\textrm{Re } z - \textrm{Re } \epsilon_{n} )^{2} + (\textrm{Im } z - \textrm{Im } \epsilon_{n} )^{2} + \eta^{2}}.
\end{equation}
We want to find an effective band structure in a disordered environment where the translational symmetry is effectively restored in a self-averaging manner. To do so, we sum the contributions from all energy levels $\epsilon_n$ with a weighting function $p(z,\epsilon_n)$ in each disorder configuration. Then, we average it over distinct disorder configurations to obtain an effective band structure. The resulting density of states would have a sharply peaked structure in the complex energy plane of $z$ as shown in Fig. \ref{fig:2} (a) and (b).

We exploited this approach to compute other quantities. To compute the order parameter spectral density in Fig. \ref{fig:2} (c), we utilized the following expression
\begin{equation}
O(z) = \frac{1}{N_{s}}\sum_{s}\frac{1}{N}\sum_{n} q_{n} p(z,\epsilon_n),
\end{equation}
where $q_n$ is given in Eq.(\ref{eq:quadrupole moment}). Lastly, to compute the local density of states in Fig. \ref{fig:2} (d), we utilized the following expression
\begin{equation}
L(\mathbf{r}) = \frac{1}{N_{s}}\sum_{s}\frac{1}{N}\sum_{n} |\phi_{n}(\mathbf{r})|^{2} p(z,\epsilon_n),
\end{equation}
where $\phi_{n}(\mathbf{r})$ is a wave function of an eigenstate $\left | n \right \rangle$.

\subsection{self-consistent Born approximation}
\subsubsection{Derivation of self-energy expression}
In this section, we calculate the disorder-averaged Green functions using the self-consistent Born approximation. The disorder-averaged Green function, $\bar{G}$ is defined as,
\bea
\bar{G} \equiv \langle G_{\textrm{dis}} \rangle=\langle \frac{1}{z-H_0-V_{\textrm{dis}}} \rangle,
\eea
where where $\langle ... \rangle$ indicates the disorder averaged quantity. $z$ is the complex energy. We can expand the Green function in the series of the disorder term as,
\begin{equation}
G_{\textrm{dis}} = \frac{1}{z-H_0-V_{\textrm{dis}}} = G_0+G_{0}V_{\textrm{dis}}G_{0}+G_{0}V_{\textrm{dis}}G_{0}V_{\textrm{dis}}G_0 + \cdots,
\end{equation}
where $G_{0}(z)=(z-H_{0})^{-1}$ is a clean Green function. When averaging over the disorder configurations, $G_{0}V_{\textrm{dis}}G_{0}$ term vanishes since $\langle w_i \rangle=0$. The lowest-order correction to the Green function is given as,
\begin{equation}
\bar{G}\approx G_0 +\langle  G_{0}V_{\textrm{dis}}G_{0}V_{\textrm{dis}}G_0\rangle.
\end{equation}
The lowest-order correction can be further evaluated in the matrix form as,
\begin{equation}
\langle  G_{0}V_{\textrm{dis}}G_{0}V_{\textrm{dis}}G_0\rangle _{ij} = \sum_{\alpha,\beta,\gamma,\delta} \langle  G_{0}V_{\textrm{dis}}G_{0}V_{\textrm{dis}}G_0\rangle _{ij} = \sum_{\alpha,\beta,\gamma,\delta,\lambda} \langle  [G_{0}]_{i\alpha}[V_{\textrm{dis}}]_{\alpha\beta}[G_{0}]_{\beta\gamma}[V_{\textrm{dis}}]_{\gamma\lambda}[G_0]_{\lambda j}\rangle,
\end{equation}
where $i,j,\alpha,..\lambda$ index indicates the real-space sites. Since $w_{i}$ in different site has no correlations, $\langle w_i w_j \rangle=\frac{W^2}{12}\delta_{ij}$, we can simplify the above expression as,
\begin{equation}
\sum_{\alpha,\beta} \langle  [G_{0}]_{i\alpha}[V_{\textrm{dis}}]_{\alpha\alpha}[G_{0}]_{\alpha\beta}[V_{\textrm{dis}}]_{\beta\beta}[G_0]_{\beta j}\rangle = \frac{W^2}{12} \sum_{\alpha,\beta}  [G_{0}]_{i\alpha}[G_{0}]_{\alpha\alpha}[G_0]_{\alpha j}.
\end{equation}
As a result we have the following correction in the disorder-averaged Green function
\bea
[\bar{G}]_{ij}\approx [G_0]_{ij} +\frac{W^2}{12}  [G_{0}]_{i\alpha}[G_{0}]_{\alpha\alpha}[G_0]_{\beta j}.
\eea
We compare the above expression with the Dyson equation, $[G]_{ij}=[G_0]_{ij}+[G_0]_{i\alpha}[\Sigma]_{\alpha\beta}[ G_0]_{\beta j} + \cdots$. We notice that the self-energy, $\Sigma$, can be written as,
\begin{equation}
[\Sigma]_{ij}=\frac{W^2}{12} \delta_{ij} [G_{0}]_{ij}.
\end{equation}
The self-energy can be re-written in the momentum space as,
\begin{equation}
\Sigma(z)=\frac{W^2}{12}\frac{1}{N}\sum_\mathbf{k} \frac{1}{z-H_0(\mathbf{k})},
\end{equation}
where $N$ is the number of the lattice sites. We may promote this first-order self-energy into a self-consistent one as
\begin{equation}
\Sigma(z) = \frac{W^2}{12}\frac{1}{N}\sum_\mathbf{k}\frac{1}{z-H_0(\mathbf{k})- \Sigma(z)}.\label{eq.born_anderson_implicit}
\end{equation}

Now, we find the symmetries that $\Sigma(z)$ should satisfy. To do so, we utilize the following symmetries of the clean Hamiltonian\cite{PhysRevB.102.205118}
\begin{subequations} \label{eq.H0_sym}
\bea
\sigma_x H_0^{T} (k_x, k_y) \sigma_x &=&  H_0(-k_x, k_y), \label{eq.H0_sym1} \\
\sigma_z H_0^{T} (k_x, k_y) \sigma_z &=& H_0(k_x, -k_y), \label{eq.H0_sym2} \\
\sigma_z H_0^{\dagger} (k_x, k_y) \sigma_z &=& - H_0(-k_x, k_y), \label{eq.H0_sym3} \\
\sigma_x H_0^{\dagger} (k_x, k_y) \sigma_x &=& - H_0(k_x, -k_y). \label{eq.H0_sym4}
\eea
\end{subequations}
Applying the transformations in Eqs.(\ref{eq.H0_sym}) to Eq.(\ref{eq.born_anderson_implicit}), we find that $\Sigma(z)$ should satisfy the following equations
\begin{subequations} \label{eq.sym_Sigma_from_H0}
\bea
\Sigma(z) &=& \sigma_x \Sigma^{T} (z) \sigma_x, \\
\Sigma(z) &=& \sigma_z \Sigma^{T} (z) \sigma_z, \\
\Sigma(z) &=& -\sigma_z \Sigma^{\dagger} (-z^*) \sigma_z, \\
\Sigma(z) &=& -\sigma_x \Sigma^{\dagger} (-z^*) \sigma_x.
\eea
\end{subequations}
If we write $\Sigma$ as $\Sigma = \Sigma_0 + \Sigma_x \sigma_x + \Sigma_y \sigma_y + \Sigma_z \sigma_z$, then $\Sigma_x$ and $\Sigma_z$ should vanish according to the above equations. Finally, we obtain the following expressions
\begin{subequations} \label{eq.born_anderson}
\bea
\Sigma_0 (z) = \frac{W^2}{12}\frac{1}{N}\sum_{\mathbf{k}}\frac{z+i(\gamma+\lambda\cos k_x)-\Sigma_0(z)}{(z+i(\gamma+\lambda\cos k_x)-\Sigma_{0}(z))^2-\lambda^2 \sin^2k_x-(\gamma+\lambda\cos k_y-\Sigma_{y}(z))^2-\lambda^2 \sin^2k_y}, \\
\Sigma_y (z)=\frac{W^2}{12}\frac{1}{N}\sum_{\mathbf{k}}\frac{(\gamma+\lambda\cos k_y)-\Sigma_y(z)}{(z+i(\gamma+\lambda\cos k_x)-\Sigma_{0}(z))^2-\lambda^2 \sin^2k_x-(\gamma+\lambda\cos k_y-\Sigma_{y}(z))^2-\lambda^2 \sin^2k_y}.
\eea
\end{subequations}
We note that $\Sigma_0$ and $\Sigma_y$ should satisfy the following identities
\begin{subequations} \label{eq.sym_Sigma_anderson}
\bea
\Sigma_0(z) & = & -\Sigma_0^* (-z^*), \\
\Sigma_y(z) & = & \Sigma_y^* (-z^*), \\
\Sigma_0(z) & = & \Sigma_0 ^* (z^*-2 i \gamma), \\
\Sigma_y(z) & = & \Sigma_y ^* (z^*-2 i \gamma),
\eea
\end{subequations}
where the first two come from Eq.(\ref{eq.sym_Sigma_from_H0}) while the latter two come from the fact that the integrand in Eq.(\ref{eq.born_anderson_implicit}) is invariant under $z \rightarrow z^ * - 2 i \gamma$ and $k_x \rightarrow k_x + \pi/2$.

We now consider the $C_{4}$-asymmetric disorder case. The difference with  the Anderson disorder case is that the disorder term has a non-trivial matrix factor of $(I_2-i\sigma_y)$. Using Eq.(\ref{eq.born_anderson_implicit}), we obtain
\begin{equation}
\Sigma(z)=\frac{W^2}{12}\frac{1}{N}\sum_\mathbf{k}(I_2-i\sigma_y)(z-H_0(\mathbf{k})- \Sigma(z))^{-1}(I_2-i\sigma_y).\label{eq.born_c4asym_implicit}
\end{equation}
Applying the transformations in Eqs.(\ref{eq.H0_sym}) to Eq.(\ref{eq.born_c4asym_implicit}), we find that $\Sigma(z)$ should satisfy Eq.(\ref{eq.sym_Sigma_c4asym}). If we write $\Sigma$ as $\Sigma = \Sigma_0 + \Sigma_x \sigma_x + \Sigma_y \sigma_y + \Sigma_z \sigma_z$, then $\Sigma_x$ and $\Sigma_z$ should vanish. Finally, we obtain the following expressions
\begin{subequations} \label{eq.born_c4asym}\bea
\Sigma_0 (z)=\frac{-i W^2}{6}\frac{1}{N}\sum_{\mathbf{k}}\frac{(\gamma+\lambda\cos k_y)-\Sigma_y(z)}{(z+i(\gamma+\lambda\cos k_x)-\Sigma_{0}(z))^2-\lambda^2 \sin^2k_x-(\gamma+\lambda\cos k_y-\Sigma_{y}(z))^2-\lambda^2 \sin^2k_y}, \\
\Sigma_y (z)=\frac{-i W^2}{6}\frac{1}{N}\sum_{\mathbf{k}}\frac{z+i(\gamma+\lambda\cos k_x)-\Sigma_0(z)}{(z+i(\gamma+\lambda\cos k_x)-\Sigma_{0}(z))^2-\lambda^2 \sin^2k_x-(\gamma+\lambda\cos k_y-\Sigma_{y}(z))^2-\lambda^2 \sin^2k_y}.
\eea\end{subequations}
We note that $\Sigma_0$ and $\Sigma_y$ should satisfy the following identities
\begin{subequations} \label{eq.sym_Sigma_c4asym}
\bea
\Sigma_0(z) & = & -\Sigma_0^* (-z^*), \label{eq.sym_Sigma_c4asym1} \\
\Sigma_y(z) & = & \Sigma_y^*(-z^*), \label{eq.sym_Sigma_c4asym2} \\
\Sigma_0(z) & = & \Sigma_0 ^* (z^*-2 i \gamma), \label{eq.sym_Sigma_c4asym3} \\
\Sigma_y(z) & = & \Sigma_y ^* (z^*-2 i \gamma), \label{eq.sym_Sigma_c4asym4}
\eea
\end{subequations}
where the first two come from Eq.(\ref{eq.sym_Sigma_from_H0}) while the latter two come from the fact that the integrand in Eq.(\ref{eq.born_c4asym_implicit}) is invariant under $z \rightarrow z^ * - 2 i \gamma$ and $k_x \rightarrow k_x + \pi/2$.

\subsubsection{Numerical solution} \label{sec.}
\begin{figure}[t!]
	\centering\includegraphics[width=1\textwidth]{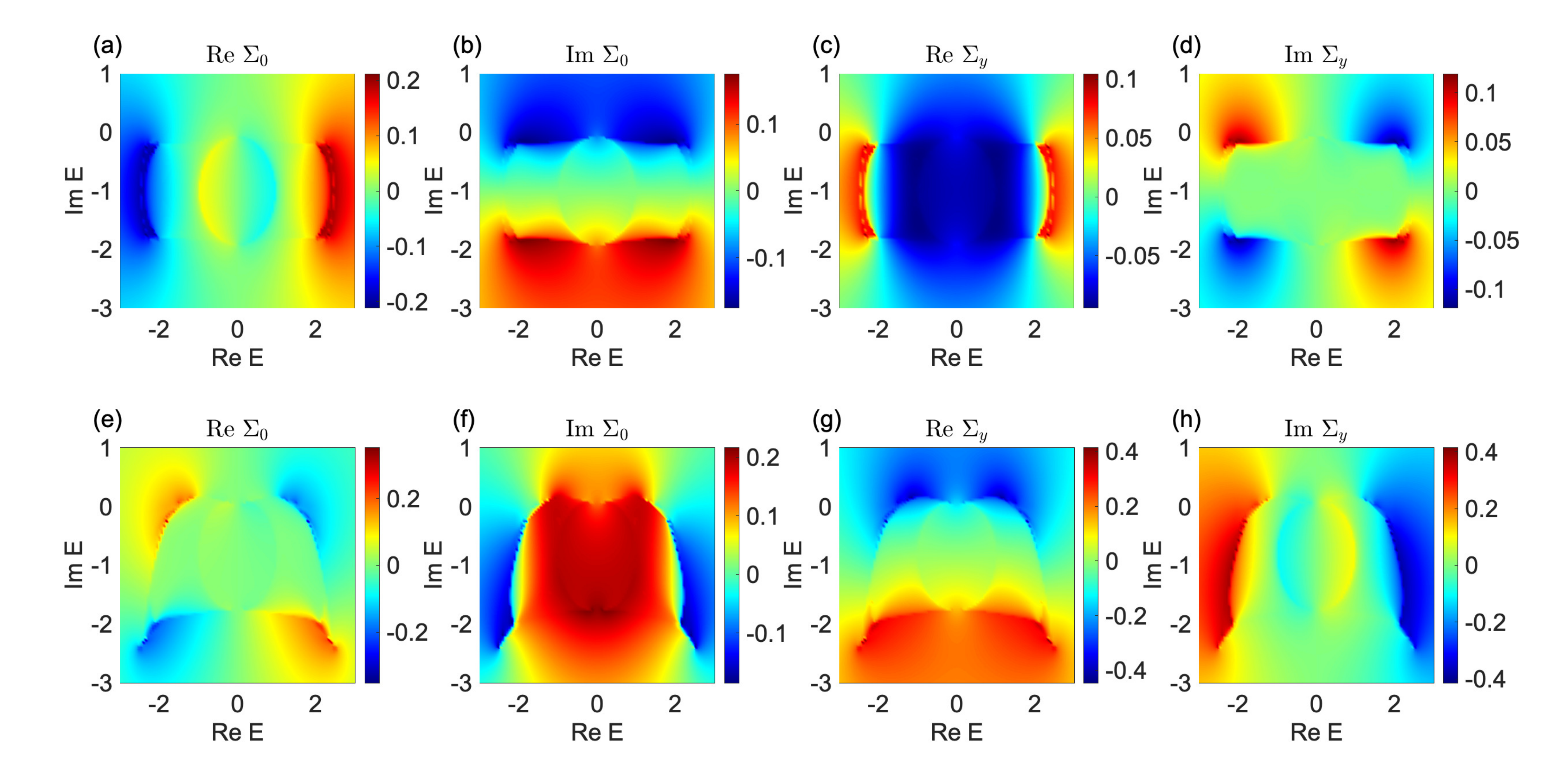}
	\caption{Numerically calculated self-energy corrections (a)-(d) for Anderson disorder  in Eq.(\ref{eq.born_anderson}) and (e)-(h) for $C_{4}$-asymmetric disorder in Eq.(\ref{eq.born_c4asym}). In all plots, $\lambda=1$, $\gamma=1$, and $W=2$ are used. }
	\label{figs2} 
\end{figure}
Here, we explain how we obtained the effective band structure in Fig. \ref{fig:2} (a) by using Eqs.(\ref{eq.born_anderson}) and (\ref{eq.born_c4asym}). In Eqs.(\ref{eq.born_anderson}) and (\ref{eq.born_c4asym}), the self-energy corrections possess an energy dependence, which turns out to be crucial to understand numerical diagonalization results. We exploited the following iterative method to solve those equations with keeping the whole energy dependence. In the first iteration, we start with the initial values of $\Sigma_0=0$ and $\Sigma_z=0$. We insert them into the right-hand sides and perform the integration to obtain $\Sigma_0$ and $\Sigma_z$. The resulting $\Sigma_0$ and $\Sigma_z$ are inserted again to obtain new $\Sigma_0$ and $\Sigma_z$. We repeat the computation until $\Sigma_0$ and $\Sigma_z$ converge to some values. We found that the iterative computation converges quite well within twenty iterations if $W$ is not too large, say $W<3$ or so. The computation should be done for each value of $z$. Figure \ref{figs2} shows the numerical results for the Anderson disorder case in (a)-(d) and for the $C_{4}$-asymmetric case in (e)-(h).

To take into account the effect of the self-energy, we define an effective Hamiltonian as follows
\begin{equation}
H_{eff}(\mathbf{k},z) \equiv H_{0}(\mathbf{k}) + \Sigma(z).
\end{equation}
By solving the eigenvalue equation for $H_{eff}$, we may obtain an effective band structure that captures the effect of disorder. Note that the equation is a non-linear one because $H_{eff}$ has an explicit dependence on $z$ that should be computed from $H_{eff}$, actually. To solve this non-linear problem, we implemented another self-consistency loop for $H_{eff}$. In the first iteration, we get the initial value of $z_0$ from $H_0$. We compute $\Sigma(z_0)$ by using the self-consistency loop for $\Sigma$ that we explained above. Then, we solve an eigenvalue equation for new $H_{eff}(\mathbf{k}, z_0) = H_0(\mathbf{k}) + \Sigma(z_0)$ to find a new $z$. We repeat the computation until $z$ converges to some value. We found that the iterative computation converges quite well within twenty iterations if $W$ is not too large, say $W<3$ or so. The computation should be done for each values of $k_x$ and $k_y$. The resulting effective band structure turns out matched quite well with the numerical diagonalization result as shown in Fig. \ref{fig:2} (a).

\end{widetext}
\end{document}